\documentstyle[epsfig]{aipproc}

\begin{document}
\title{Spectral Modeling of GRB Pulses}

\author{Hara Papathanassiou}
\address{International School for Advanced Studies (SISSA), Trieste, Italy}
%\lefthead{LEFT head}
%\righthead{RIGHT head}
\maketitle

\begin{abstract}
The energy spectra of pulses of GRBs are modeled 
for synchrotron  and multiple self-inverse Compton scatterings from 
a population of  thermal and non-thermal $e^-$s. The contribution 
from  pairs that result from annihilation is also taken into account.
A high particle density (enhanced by the pairs) will
increase absorption but, if the pairs are not accelerated,
the absorption frequency cannot lie in the BATSE window.
Pairs will contribute in upscattering and will most likely
increase the population of cold  particles that will
downscatter hard photons and thus suppress 
hard  (e.g., above 300~keV) emission.

\end{abstract}

\section*{Introduction}
The properties  of time resolved  BATSE spectra
can probe the actual physical mechanisms responsible for
the emission in this range. Time integrated spectra, on
the other hand, result from the integration of time-dependent quantities which tends to smooth out
 spectral features. In the case of GRBs, such an averaging is unavoidable due to the
photon time delay and aberration inherent in a relativistically moving medium.
Spectral evolution sequences or average spectra over the course of the whole burst will 
 reflect the hydrodynamical evolution of the quantities
involved in the radiation processes.
Quantities like the bulk Lorenz factor $\Gamma_b$ of the
flow, or the available energy, will remain roughly
constant during the generation of a pulse. Therefore, spectral modeling of pulses addresses best the radiative mechanisms.
Spectral fits of a large number of bright time-resolved spectra \cite{Preece99}
have furnished distributions of the three main spectral
parameters: the low ($\alpha$) and high  ($\beta$) photon spectral indices and the frequency of the spectral 
break ($E_b$). These quantities are primarily linked with the emission processes.
The surprising narrow $E_b$ distribution probably holds  the key answer to the physical mechanisms (be them the radiation process itself, or the mechanism for magnetic field amplification and particle acceleration). In the widely used synchrotron self- inverse Compton model for GRB spectra, $E_b$ is a function of free parameters
and can only be used to constrain them \cite{mythesis}.

The low energy spectral index $\alpha$ has a distribution \cite{Preece99} that can be approximated by a Gaussian centered at $\alpha = -1$, with a FWHM of $\sim 0.9$ plus a separate component of a small percentage of cases with $\alpha = 1$. The radiative mechanisms that would {\it intrinsically} result in these values are: Comptonization by moderate optical depth \cite{Ghis&Cel}, 
or bremsstrahlung in $\alpha \approx -1$,
optically thin synchrotron from a (cooled) electron power law in $\alpha = -2/3 (-3/2)$, inverse Compton (IC) in $\alpha =0$, optically thick emission by any mechanism in the Rayleigh-Jeans limit in $\alpha = 1$. More importantly, any 
steep intrinsic spectral slope portion of the spectrum
will be observed  flatter
 due to sampling over volume, energy, time distributions (e.g.,
a self absorbed synchrotron spectrum with the
self-absorption frequency, $\nu_{abs}$,  sweeping
through the window of interest due to an increase in the thickness of the medium will result in a photon index of 1, rather than 1.5 \cite{myroma}).
Further inferred flattening of the value of $\alpha$ is
introduced when fitting a spectrum through model
broadband  functions because, in practice, the BATSE
window is too narrow \cite{Preece99}, \cite{Nicole}.
The large number of cases with $\alpha >-2/3 $,  coined as the ``death line
of the synchrotron model'' \cite{Preece98}, suggests  that any model employed will not be in
a fully optically thin regime in the BATSE window.

The high energy power law index $\beta$ provides the  best indication for the 
acceleration of particles to a power law (of index
$p$). The distribution of the fitted values in the time
resolved sample \cite{Preece99} is bimodal, its main
component peaking at $\beta \approx-2.25$. If interpreted
as synchrotron emission from particles that have cooled
down, it gives $p=2.5$. This is in agreement with the values derived from afterglows
and is reproduced by particle acceleration calculations \cite{Yve}.
Attributing the full range of $\beta$ values to
synchrotron emission from a power law, suggests that the
radiation populations of $e^-$ have $2\le p \le 4$ -but
invoking IC could narrow this range down to $2 \le p \le 3$.
Roughly 10\% of the spectra have no high-energy power law
component. These spectra can be either those classified
as no high energy (NHE) pulses \cite{HE_NHE} or those with $\beta \le
-4$. Such values can be interpreted as representing the
Wien part of spectrum (e.g., resulting from  a thermal particle
distribution) or a power law spectrum that has suffered some absorption at the high end.

Here, I report on spectral evolution sequences for GRB
 pulses. I discuss the spectra in the comoving frame
 only,
 since the purpose is to account for spectral slopes that
 are sufficiently steep, so that they can result in
 observed spectra consistent with the fits to  BATSE data (e.g., \cite{Preece99}).

\section*{SYNCHROTRON SELF ABSORPTION}
\begin{figure}[b!] % fig 1
\centerline{\epsfig{file=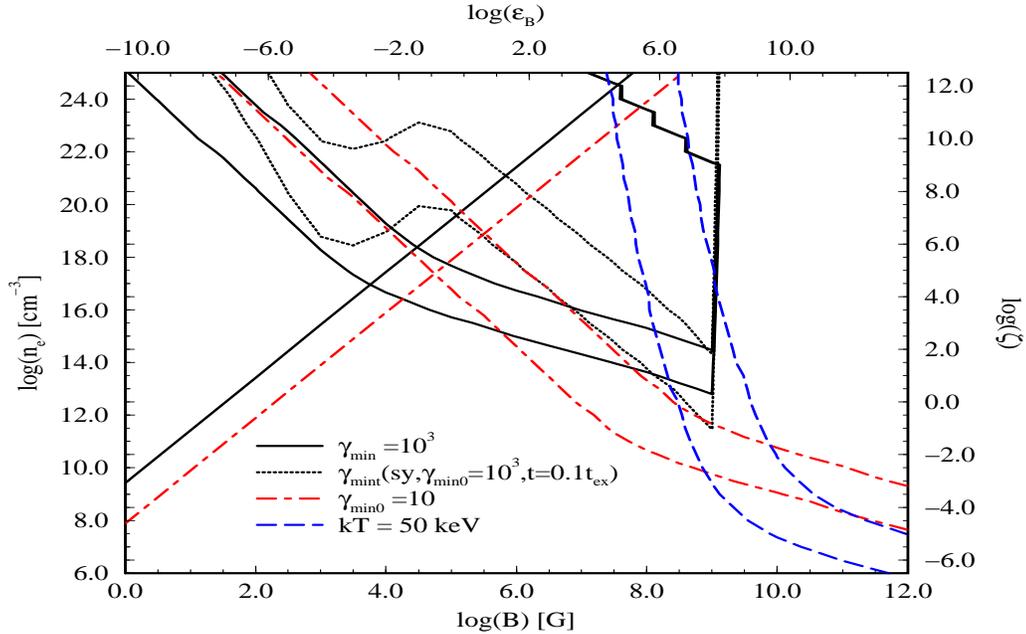,angle=0,height=3.5in,width=5.5in}}
\vspace{10pt}
\caption{Values of the magnetic field and the particle
number density for which the synchrotron spectrum turns
optically thick in the 30-300 keV range. On the facing
axes are given the corresponding equipartition values for
a flow of $t_{var} =0.01$~s, $L_{52} =1$, and $\Gamma_{b}
= 300$. }
\label{fig:v_abs}
\end{figure}
As long as a {\it{high}} $\alpha$ spectrum is produced locally, all lower $\alpha$ values can be reproduced by the integrations necessary to obtain the observed spectra and can be attributed to time dilation and aberration of photons,
adiabatic dilution of the $e^-$ distribution, inhomogeneous conditions in the emitting region.
Here, I examine whether a steep slope spectrum ($\alpha \ge 1$) can be generated in an internal shock flow (external shocks are produced further out in the flow and thus involve less dense material). I pick typical values for the global parameters: total available luminosity per unit solid angle of $10^{52}$~erg/s/sterad ($L_{52}/\Omega =1$), variability timescale $t_{var} = 0.01$~s, and  $\Gamma_b =300$ (note that $\Gamma_b$ cannot be arbitrarily large even for a very clean flow because of Compton drag, \cite{MR_internal}).
Figure \ref{fig:v_abs}  shows the range in particle
number density $n_e$ and magnetic field $B$ that would
place $\nu_{abs}$  in the  30-300 keV range. This
frequency is defined as the one where the optical depth
to synchrotron is 1. The spectral slope below that will
be 1 (or 1.5 for emission from a power law that is
optically thick above the peak).
 The calculation is performed for a power of $e^-$s with
index $p=2.5$ (the results depend very weakly on $p$)
peaking at $\gamma_{min0} =10^3$, one peaking at
$\gamma_{min0} =10$, and a thermal distribution with $kT
=50 $~keV. These constitute the simplest cases one would
test.
A more realistic case, where the $e^-$  spectrum  results
from continuous ejection with constant density and
cooling via synchrotron emission only, for the first
1/10th of the duration of the pulse, is calculated as well. To the left of the diagonal lines, the dominant cooling mechanism for the $e^-$s at peak energy is bremsstrahlung (therefore, the usual
assumption of the emission being due synchrotron is no
longer valid). 
As seen, in order to get $\nu_{abs}$ in the BATSE range,
for an $e^-$ distribution peaking at $\gamma_{min0}
=10^3$ and for an equipartition magnetic field value, one
would need the number density  of particles to be roughly
$10^3$ to $10^5$  of that of the protons in the flow. If one allows for the cooling of the $e^-$ distribution, the requirement on $n_e$ is more severe. 
Having a thermal particle distribution, while it relaxes the constraint on $n_e$ it requires a flow that is strongly magnetically dominated. 
Having a shorter variability timescale (and/or lower $\Gamma_b$) 
implies a denser environment and this results in lower equipartition fractions
(e.g., for $t_{var} = 1$~ms and $\Gamma_{b} = 200$, $\epsilon_{B} \approx 1$ would require $\zeta =10^3$) but also makes bremsstrahlung the dominant cooling mechanism for the $e^-$ peak.
One could therefore conclude that it is not possible to have $\nu_{abs}$ in the BATSE window, if the emitting environment is that of internal shocks and the emission mechanism is synchrotron of the available $e^-$s.
There is a possibility though to substantially increase the number density of
emitting particles by including the $e^- e^+$ pairs that are produced in the flow due to pair opacity of the interactions of hard IC photons \cite{myroma}, \cite{Ghis&Cel}.
But these pairs should be added with high densities and very hard spectra (i.e., be produced abundantly and get accelerated immediately).

\section*{SPECTRAL EVOLUTION SEQUENCES}
I calculate spectral evolution sequences for continuous
injection of particles with a prescribed power distribution
(consisting of a relativistic Maxwellian at low energies,
$\sim \gamma^2$,  and a power law, $\sim \gamma^{-p}$
above the peak, $\gamma_{min0}$, and constant number density)
and losses through adiabatic expansion, synchrotron and
IC radiation
in a region that is expanding at constant rate. The
synchrotron and multiply (up to $\sim 10$) IC upscattered spectral
components that result form the distribution at hand are
calculated.
The pair production is evaluated following the
prescription of \cite{Bonometto_Rees} (valid for 
scattering of photons with very different energies). This
allows us to evaluate the attenuated hard spectrum, the
pair distribution (which is a power law peaking at $\gamma
\approx 1$  with index that of the
hard photons).
At each timestep, the number of $e^-$s with energies
in the first bin are added to a population of cold
$e^-$s which are used in calculating the down-scattered
spectrum.
At the end of each cycle the pairs are added to the
$e^-$ distribution at an average constant rate (following two different
prescriptions:
({\it i}) the aforementioned power law, ({\it ii}) a
thermal distribution of 50 keV \cite{Ghis&Cel}).
Details of the calculation will appear in a forthcoming
paper.

Values of parameters close to equipartition and a low
$\Gamma_b$
turn the spectra optically thick to pair production. In
the absence of reacceleration, the pairs will contribute
to absorption ($\nu_{abs}$
can reach up to a few keV), modest upscattering, and will
provide abundant cold particles that down-scatter hard
photons.
Given a sufficient number density of cold $e^-$s in the flow, the hard photons
will lose energy in successive scatterings.
If the optical depth of cold $e^-$s is $\tau_c$, and for
scatterings in the Thompson regime, a cut-off will appear
at around $\Gamma_b m_e c^2 /\tau_c^2$.
 Requiring such a cutoff at 300~keV introduces a lower limit on the number of cold $e^-$s (and $e^+$) in the flow of $n_c \ge 6.5\, 10^{13} /\sqrt{\Gamma_b} t_{var}$. Even if all the $e^-$s of the flow (those coming from the
ionization of the explosion material) were cold they
would not be able to account for the cut-off. This is
possible in a situation where the flow is very optically thick to pairs. Pair production interactions of photons
eliminate all  photons  that in the comoving frame have
energies above 0.5 MeV. This creates a large number of
cold pairs that subsequently down-scatter the hard
photons (all scatterings take place in the Thompson
regime).
 A rough estimate shows that for a cut-off
at 300~keV, the product of the $e^-$ and IC radiative
efficiency must be 
$\varepsilon_e \varepsilon_{ic} \approx 0.077 \frac{\eta_2 \Gamma_{b2}^{9/2} t_{var}}{L_{52}/\Omega}$
where  the specific entropy of the flow $\eta$ and $\Gamma_b$
are measured in units of 100.

The $\beta$ distribution argues in favor of a particle acceleration mechanism
resulting to a power law distribution in most of the pulses.
A simple way (and an alternative one to the downscattering
by cold $e^-$s) to explain the very steep high energy
spectra (or NHE pulses, \cite{HE_NHE}) 
is by invoking a 
failed particle acceleration process. In this case, the $e^-$ distribution is 
either thermal, or a narrow Gaussian centered at
$\gamma_{min0}$ and the power law extension is suppressed.

\section*{CONCLUSIONS}
Observed synchrotron spectra can have {\it any} $\alpha$ value above -2/3.
The $\sim 30\%$ of the spectra that require a higher value imply optically thick conditions in the BATSE window. For this, one infers high values of particle densities and a consistent spectral modeling calls for inclusion of bremsstrahlung emission.
Pairs are present in the flow and they reach maximum
density at
around the pulse peak. They contribute to absorption, but
since they are created with low energies they cannot push
$\nu_{abs}$ into the BATSE window, {\it unless} they are
accelerated. They may cause a changing spectral index
through Comptonization (for very short pulses).
They can provide a time varying population of cold particles that may be responsible for the lack of high energy emission (or steep fall off) of about $10\%$ of the spectra.

\end{document}